\documentclass[runningheads,envcountsame,envcountsect,oribibl]{llncs}

\usepackage[latin1]{inputenc}
\usepackage[colorlinks=true,ps2pdf=true]{hyperref}
\usepackage[dvips]{graphics}
\usepackage{psfrag}
\usepackage{amsmath}
\usepackage{amssymb}
\usepackage[english]{babel}
\usepackage{xspace}
\usepackage{url}

\def\rcs$#1${#1}

\newcommand{\eg}{{e.g.},\ }		
\newcommand{\etal}{\textit{et al.}~}	
\newcommand{\ie}{{i.e.},\ }		
\newcommand{\cf}{{cf.}\ }               
\newcommand{\id}[1]{\ensuremath{\mathit{#1}}}	
\newcommand{\setjoin}{\cap}
\newcommand{\card}[1]{\lvert{#1}\rvert}		
\renewcommand{\emptyset}{\varnothing}	

\newcommand{\Prob}[1]{\mathord{\textbf{Pr}}\left[#1\right]} 

\newcommand{\cset}{B}
\newcommand{\csset}{B}
\newcommand{\scset}{\mathbf{B}}
\newcommand{\cspace}{\mathbf{B}}
\newcommand{\cssize}{b}
\newcommand{\cspacesize}{\beta}

\newcommand{\faulty}{f}
\newcommand{\corrupted}{d}
\newcommand{\dlimit}{r}
\newcommand{\cid}{\id{cid}}
\newcommand{\sig}[2]{[#1]_{#2}}
\newcommand{\coin}{c}
\newcommand{\secpar}{s}  

\makeatletter
\newcommand{\xRightarrow}[2][]{\ext@arrow 0359\Rightarrowfill@{#1}{#2}}
\newcommand{\xLeftarrow}[2][]{\ext@arrow 3095\Leftarrowfill@{#1}{#2}}
\newcommand{\@Setstar}[1]{\left\{{#1}\right\}}
\newcommand{\@Set}[2]{\@Setstar{{#1},\ldots,{#2}}}
\newcommand{\Set}{\@ifstar{\@Setstar}{\@Set}}	
\makeatother

\newcommand{\sendright}[1]{$\xrightarrow{\quad #1\quad}$}
\newcommand{\sendleft}[1]{$\xleftarrow{\quad #1\quad}$}

\title{Distributed Double Spending Prevention\thanks{%
This research is/was partially supported by the research program 
Sentinels (\url{www.sentinels.nl}), project JASON (NIT.6677).
Sentinels is being financed by Technology Foundation STW, the Netherlands Organization
for Scientific Research (NWO), and the Dutch Ministry of Economic Affairs.\hfil\break
\rcs$Id: double-spending.tex 18 2008-02-06 14:01:34Z jhh $}}

\author{Jaap-Henk Hoepman} 
\institute{TNO Information and Communication Technology\\
  P.O. Box 1416, 9701 BK \ Groningen, The Netherlands\\
  \email{jaap-henk.hoepman@tno.nl}\\
  and\\
  Institute for Computing and Information Sciences \\
  Radboud University Nijmegen\\
  P.O. Box 9010, 6500 GL \ Nijmegen, the Netherlands\\ 
  \email{jhh@cs.ru.nl}}
\begin{document}

\maketitle

\bibliographystyle{alphacm-}

\begin{abstract}
We study the problem of \emph{preventing} double spending in electronic payment
schemes in a \emph{distributed} fashion. 
This problem occurs, for instance, when the spending of electronic coins
needs to be controlled by a large
collection of nodes (\eg in a peer-to-peer (P2P) system) instead of one central
bank. 
Contrary to the commonly held belief that this is fundamentally impossible,
we propose several solutions that do achieve a reasonable level of
double spending prevention, and 
analyse their efficiency under varying assumptions.
\end{abstract}

\section{Introduction}

Many electronic payment schemes exist.
For an overview, we refer to Asokan~\etal\cite{asokan1997state-of-art-payment}
or O'Mahony~\etal\cite{MahPT97}.
Some of those are coin based, where some bitstring locally stored by a user
represents a certain fixed value. 

Coin based systems run the 
risk that many copies of the same bitstring are spent at different merchants.
Therefore, these systems need to incorporate \emph{double spending} prevention
or detection techniques. 
To \emph{prevent} double spending, a central bank is usually
assumed which is involved in each and every transaction. 
In off-line scenarios
(where such a connection to a central bank is not available), double spending
\emph{detection} techniques are used that will discover double spending at
some later time, and that allow one to find the perpetrator of this illegal
activity. 
A major drawback of double spending detection techniques is the risk
that a dishonest user spends a single coin a million times in a short period of
time before being detected. 
This is especially a problem if such a user cannot be punished for such
behaviour afterwards, \eg fined, penalised judicially, or being kicked from the
system permanently.

Recently, the use of electronic payment like systems has been
proposed\footnote{%
  America Online and Yahoo announce introduction of electronic postage
  for email messages ("Postage is Due for Companies Sending E-Mail", New York
  Times, February 5, 2006).}
to
counter SPAM~\cite{hird2002technicalcontrolspam} or to enforce fairness among users of peer-to-peer (P2P)
networks~\cite{Yang+Garcia-Molina-PPay:03,Vishnumurthy+ETAL-KARMA:03,garcia2005karma}.
In such systems it is unreasonable to assume a central bank, either  because it
does not exist, or because it would go against the design philosophy of the
system (as is the case for P2P networks). 
At first sight it then appears to be impossible to prevent double spending.
This would limit the usefulness of such approaches because of the rapid double
spending problem described above: users can easily rejoin a P2P system under a
different alias and continue their bad practises forever.

In~\cite{garcia2005karma} 
we wrote:
%
\begin{quote}
We note that for any system offering off-line currency, double-spending
\emph{prevention} is generally speaking not possible, unless extra assumptions
(\eg  special tamper proof hardware) are made.
\end{quote}
In that paper, in fact, we were not considering a completely off-line system,
but a decentralised system without a central bank instead. 
The difference turns out to be decisive. In a truly off-line system
(where the receiver of a coin has no network access to perform any kind of
checking, and where the spender of a coin is not forced to adhere to a security
policy
 through some kind of tamper proof hardware~\cite{SchS99}) the chances of double spending
prevention are slim. We soon after realised, however, that
the situation is not so bad in
an on-line but decentralised system without a central bank.

The crucial observation is that it may be impossible, or very expensive, to
prevent every possible double spending of a coin (\ie a deterministic
approach), but that it may very well be possible to prevent that
a particular coin is double spent \emph{many times}, using efficient randomised
techniques.
Even such a weaker guarantee limits the damage an adversary can do.
In other words, the main paradigm shift is the realisation that
double spending a single coin twice is not so bad, but
spending it a hundred times should be impossible.
Of course, such a probabilistic and limited security property may not
be strong enough for the protection of `real' money. 
It may, however, be quite workable for currencies used to enforce fairness among
P2P users.

In this paper we study several such techniques for distributed double spending
prevention. We focus in this paper on methods to distribute the tasks of
the central bank over (a subset of) the nodes in the system.
An extreme case would be the distribution of the central bank over
all nodes in the system, making everyone a clerk working for the bank.
This would lead to an enormous communication overhead, as all $n$ nodes
in the system would have to be contacted for each and every transaction.
We study techniques to reduce the size of such clerk sets, mainly in
probabilistic ways, while still keeping reasonable double-spending prevention
guarantees. 

Next to a deterministic approach, there are two fundamentally different ways to
construct the clerk sets in a probabilistic manner. The most efficient method
--- yielding the smallest clerk sets --- uses the unqiue identifier of a coin
to limit the possible members of the clerk set in advance. In this model,
certain clerks attract certain coins, making it far more likely that double
spending is detected. The drawback is that given a particular coin these clerks
are known beforehand. This means the adversary has advance knowledge regarding
the clerks that it needs to bribe in order to be able to double spend a
particular coin. In certain situations this may be undesirable. Therefore we
also study the less efficient case where the clerks are selected uniformly at
random.

\subsection{Our results}

We prove the following results, where $n$ is the total number of nodes,
$\faulty$ is the total number of dishonest nodes, 
$\corrupted$ is the number of dishonest
nodes that may be corrupted by the adversary after they join the network, and
$\secpar$ is the security parameter 
(see Section~\ref{sec-model} for details).

Deterministic double spending prevention can be achieved with clerk sets
of size $2 \sqrt{n(\faulty+1)}$. 

Using randomisation double spending can be prevented with clerk sets of size
at least $\sqrt{\frac{n\secpar}{\log e (1-\faulty/n)}}$.
If we require that double spending only needs to be detected
when a single coin is double spent at least $\dlimit$ times\footnote{%
   $\dlimit$ denotes the number of times a coin is double spent.
   To be precise, when a node spends the same coin $x$ times, then $\dlimit=x-1$.
}
we need clerk sets of size at least
$\frac{\sqrt{2n\secpar}}{\dlimit}$
when $\faulty=1$ (\ie if only the double-spender itself is dishonest)
and $\sqrt{\frac{n\secpar}{\log e (1-\faulty/n)\dlimit}}$,
when $\faulty>1$.
Note that it is indeed interesting to consider the $f=1$ case seperately,
because it corresponds to the situation where nodes in the clerk sets have no
incentive to collaborate with the double spender to let him get away 
undetected, and is closely related to the selfish but rational
models used in game theoretic analysis of security protocols
(\cf~\cite{izmalkov2005rationalsecure}).

Finally
we prove that making use of the coin identifier to construct
coin specific clerk spaces of size $\cspacesize$ at least
$\corrupted + \frac{\secpar}{\log((n-\corrupted)/(\faulty-\corrupted))}$
clerk sets sampled from this space
of size at least
$\frac{\cspacesize}{\dlimit \log e} \left( \secpar + 1 + \log (\dlimit+2) \right)$
suffice to detect a coin that is double spent at least $\dlimit$ times.

These results tell us the following. Deterministically, clerk sets that have
$\sqrt{n\faulty}$ nodes suffice. For any reasonable $\faulty$  this is 
unworkable.
Using randomisation, $\sqrt{n / (1-\faulty/n)}$ is good enough.
For decent fractions of
faulty nodes (\eg $\faulty/n = 1/2$) this stays $O(\sqrt{n})$.
When we relax the double spending detection requirement and allow upto $\dlimit$
double spendings to be undetected, clerk sets can be further reduced by
a $\sqrt{\dlimit}$ factor.
Finally, if we use information stored in the coin, the size of the clerk sets
becomes independent of the size of the network, depending only on the inverse
ratio $n/\faulty$ of faulty nodes, and the number of corruptable nodes
$\corrupted$.

\subsection{Related research}

The deterministic variant of distributed double spending prevention, \ie the
one where double spending is \emph{always} prevented, is equivalent to the
problem of distributing a database over $n$ nodes, $f$ of which may be faulty.
Quorum systems (cf. \cite{malkhi1998bqs,malkhi2001pqs}) have been studied as an abstraction
of this problem, to increasing the availability and efficiency of replicated
data. A quorum system is a set of subsets (called quorums)
of servers such that every 
two subsets intersect. This intersection property guarantees that if a 
write-operation is performed at one quorum, and later a read-operation 
is performed
at another quorum, then there is some server that observes both operations and
therefore is able to provide the up-to-date value to the reader.
The clerk sets in our work
correspond to the quorums in that line of research.
We do note however that the relaxation of allowing upto $\dlimit$
double spendings to occur is not covered by the work on quorum systems.

Our approach is in a sense a dual to the one advocated by Jarecki and
Od\-lyz\-ko~\cite{jarecki1997micropayment-polling} (and similarly by
Yacobi~\cite{yacobi1998partialaudit}), in which double spending is prevented
probabilistically and efficiently by checking a payment with the \emph{central}
bank only with some probability (instead of always).

\subsection{Structure of the paper}

The paper is organised as follows. We first describe the model and the
basic system architecture in Section~\ref{sec-model}.
This fixes the way coins are represented and spent among nodes, and describes
how clerk sets are used to detect double spending. This architecture is
independent of how the clerk sets are constructed. Different construction
methods yield different performance, as described in the sections
following. It is exactly these combinatorial constructions that are the main
contributions of this paper.

We analyse the performance of fixed clerk sets in
Section~\ref{sec-fixed}, 
followed by the analysis of randomly chosen clerk sets in
Section~\ref{sec-random}.
Next, in Section~\ref{sec-multispend}, we study what happens if we allow coins
to be double spend more often, up to a certain limit $\dlimit$.
Then, in section~\ref{sec-reducing} we discuss ways to further reduce the
size of the clerk sets by making use of information in the coin.
We conclude with a thorough discussion of our results in Sect.~\ref{sec-concl}.

%
%
%
%
%
%

\section{Model and notation}
\label{sec-model}

%
We assume a distributed system consisting of $n$ nodes, at most $\faulty$ of which
are dishonest. The dishonest nodes are under the control of the adversary.
If the system is a peer-to-peer (P2P) overlay network, the nodes receive a
random identifier when joining. This identifier is not under the control of the
adversary. The adversary may, however, be able to compromise $\corrupted$ 
out of the $\faulty$
dishonest nodes \emph{after} joining the network, \ie it may compromise at most
$\corrupted$ nodes for which it knows the P2P identifier\footnote{%
  This distinction between $\faulty$ and $\corrupted$ turns out to be only
  significant in the case where coin identifiers are used to restrict the size
  of the clerk sets.
}.

Each node owns a pair of public and private keys. A signature $\sig{m}{i}$
of node $i$ on a message $m$ can be verified by all other nodes.
We let $\log$ denote the logarithm base 2.

The system handles coins, that are uniquely identified by a coin identifier
$\cid$. Valid coin identifiers cannot 'easily' be generated by nodes
themselves. Nodes can distinguish valid coins from invalid ones.
A detailed discussion on how nodes initially obtain such coins lies
outside the scope of this paper. But to argue the viability of our approach, we
briefly mention the following two options.
Coins could, for instance, be distributed
initially by a central authority. In this case, the coin identifier
incorporates a digital signature from this authority. Or they could be
generated by the nodes themselves by finding collisions in a hash function $h$
(\cf\cite{garcia2005karma}). Then, the coin identifier contains the pair
$x,y$ such that $h(x)=h(y)$. 

Nodes communicate by exchanging messages. We assume a completely connected 
network, or a suitable routing overlay. The network is asynchronous. In
particular, coins may be spent concurrently.
The network is static: no nodes join or leave the network once the system
runs.

All in all these are quite strong assumptions (a static network, with a network
wide PKI, and a point-to-point communication substrate), but not unreasonably
so. In any case, they allow us to focus on the main research issue: the
combinatorial analysis of distributing the task of an otherwise centralised
bank over the nodes of a distributed system, such that double spending is
prevented. 

The adversary tries to double spend a single coin at least $\dlimit$ times (when a
node
spends a single coin $x$ times, then $\dlimit=x-1$).
We say the system is secure with security parameter $\secpar$ if the adversary must
perform an expected $O(2^\secpar)$ amount of work in order to be successful. 
We show this
by proving that the probability of
success for the adversary for a single try is at most $2^{-\secpar}$.

We note that we do not consider denial of service attacks, for example attacks
where the clerk sets receive polluted information from dishonest nodes to
invalidate coins held by honest nodes.

\subsection{Distributing the bank}
\label{ssec-distributing}

Throughout the paper we assume the following system architecture to distribute
the bank over the nodes in the network.

\newcommand{\impref}{\rightarrow}
\newcommand{\pref}{\Rightarrow}

\begin{figure}
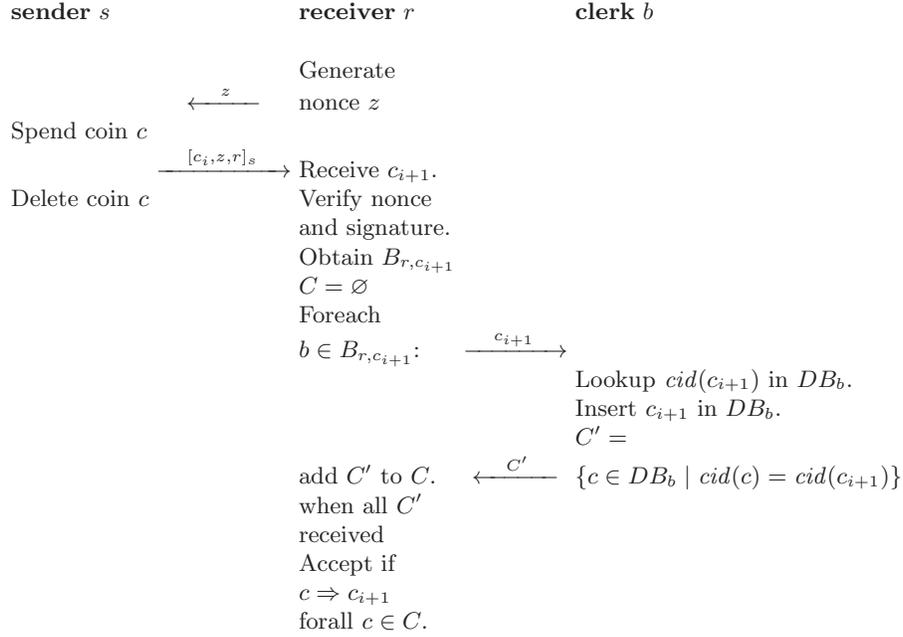

\footnotesize
\begin{center}
\begin{tabular}{lclcl}
\textbf{sender} $s$ & & \textbf{receiver} $r$ & & \textbf{clerk} $b$ \\
   & \\
   &              & Generate \\
   & \sendleft{z} & nonce $z$ \\
Spend coin $\coin$\\
   & \sendright{\sig{\coin_i,z,r}{s}} 
                  & Receive $\coin_{i+1}$. \\
Delete coin $\coin$
   &              & Verify nonce \\
   &              & and signature. \\
   &              & Obtain $\cset_{r,\coin_{i+1}}$ \\
   &	          & $C=\emptyset$ \\
   &              & Foreach\\
   &              & $b \in \cset_{r,\coin_{i+1}}$: 
			 & \sendright{\coin_{i+1}} \\ 
   &              &    &                & Lookup $\cid(\coin_{i+1})$ in $DB_b$. \\         
   &              &    &		& Insert $\coin_{i+1}$ in $DB_b$. \\
   &		  &    &                & $C'=$\\
   &              & add $C'$ to $C$.
                       & \sendleft{ C'} & $\Set*{\coin \in DB_b \mid \cid(c) = \cid(\coin_{i+1})}$  \\
   &		  & when all $C'$ \\
   &              & received\\
   & 	          & Accept if\\
   &                  & $\coin \pref \coin_{i+1}$ \\
   &              & forall $\coin \in C$.\\
    
\end{tabular}
\end{center}
\caption{Coin spending and detection protocol.}
\label{fig-prot}
\end{figure}

A coin is uniquely determined by its coin-id $\cid$. Spending a coin $\coin_i$
transfers ownership of that coin from a sender $s$ to a receiver $r$. We use
the following method (also depicted in Figure~\ref{fig-prot}): the receiver
sends a nonce $z$ to 
the sender, who then signs the coin, together with the nonce and the
name of the receiver,
sending the result 
\[
\coin_{i+1} = \sig{\coin_i,z,r}{s}
\]
back to the receiver. We call $\coin_i$ the immediate prefix of $\coin_{i+1}$ 
(denoted
$\coin_i \impref \coin_{i+1}$), and require that $s$ equals the receiver of 
$\coin_i$
(otherwise $\coin_i$ should not have been in the posession of $s$ in the first
place).  
An unspent coin simply corresponds to its coin-id $\cid$. 
$\coin$ is a prefix of $\coin'$, denoted
$\coin \pref \coin'$ if there is a sequence of coins 
$\coin_0,\ldots,\coin_k$, $k>0$ such that 
$\coin=\coin_0$, $\coin_k=\coin'$ and 
$\coin_i \impref \coin_{i+1}$ for all $0 \le i < k$.
The coin-id $\cid(\coin)$ of a coin equals its shortest prefix, or $\coin$
itself
if no prefix exists.

So called \emph{clerk sets} are used to verify the validity of a coin.
These clerk sets consist of nodes in the network that simulate a bank 
in a distributed fashion. 
The selection of nodes that are member of a clerk set $\cset_{r,\coin}$
can be either done deterministically or randomly, and may depend on both the
node $r$ accepting the coin and the coin identifier $\cid(\coin)$ of the coin
being accepted.
To perform their duties, the nodes in a clerk set
store the history of coins. 
When a receiver $r$ receives a coin $\coin$, it first verifies the signature, 
the nonce,
and the sender. It then requests from each clerk in the clerk set $\cset_{r,\coin}$
all coins with coin-id $\cid(\coin)$ that it stores.
At the same time, the clerks store $\coin$. These two steps are one atomic
operation. 
If all coins $r$ receives from its clerk set are proper prefixes of $\coin$, 
it accepts the coin. Otherwise it rejects the coin.

We note that the size of a coin increases every time it is spent, because
of the signature that must be added. Similarly, the set of coins stored by the
clerk sets grows without bounds. Dealing with these unbounded space
requirements falls outside the scope of this paper. We discuss some ways to
bound the space requirements in Sect.~\ref{sec-concl}.

The remiander of this paper assumes the above protocol for spending a coin,
and is merely concerned with different methods for
obtaining $\cset_{r,\coin_{i+1}}$ such that double spending is prevented.
The following property of the system described above is the basis for the main
results of this paper.
%
\begin{property}
Let $j$ and $k$ be honest nodes, and let $\coin$ be a coin. If
$\cset_{j,\coin} \setjoin \cset_{k,\coin}$ contains at least one honest node,
then no node can double spend a coin with coin-id $\cid(\coin)$ at both 
$j$ and $k$ using the protocol described above.
\end{property}
\begin{proof}
Let $x$ be the honest node in $\cset_{j,\coin} \setjoin \cset_{k,\coin}$.
If $i$ manages to double spend $\coin$ at both $j$ and $k$ ($j=k$ is possible),
$x$ receives a request to lookup (and immediately store) 
$\coin_j = \sig{\coin',z_j,j}{i}$ from $j$ and
$\coin_k = \sig{\coin'',z_k,k}{i}$ from $k$ (with unique nonces $z_j$ and $z_k$)
where $\cid(\coin_j)=\cid(\coin_k)$, $\coin_j \not\pref \coin_k$ and
$\coin_k \not\pref \coin_j$ (by definition of double spending).
W.l.o.g. assume $j$ makes that request to $x$ first.
Then $j$ stores $\coin_j$ at $DB_x$
before $k$ requests all coins with $\cid(\coin)=\cid(\coin_k)$.
Then $k$ retrieves $\coin_j$ with $\coin_j \not\pref \coin_k$ and hence
$k$ does not accept $\coin_k$.
\qed
\end{proof}
Observe that the inclusion of nonces in the coin spending phase is really only
necessary to determine the exact node that double-spent the coin first. 

\section{Fixed clerk sets: deterministic case}
\label{sec-fixed}

We will now study several methods to assign clerk sets to nodes. We start with
the deterministic case where each node is given a fixed clerk set $\cset_i$.
We assume $\corrupted=\faulty$ (in the deterministic case it makes no
difference whether the adversary can corrupt the nodes after they join the
network or only before that: it can ensure \emph{in advance}
to only double spend
at nodes for which the clerk sets contain no honest nodes).

If, except for the node trying to double spend,
there are no dishonest nodes, we only need to require 
$\cset_i \setjoin \cset_j \neq \emptyset$ (and the double spender should not be
the only node in that intersection). 
Clearly, we can set $\cset_i = \Set*{b}$ for all $i$ and some clerk $b$. 
This coincides with the `central bank' case described in the introduction. 
In this paper we are of course interested in the distributed case, where there should be no
single point of failure, and where the load for preventing double spending is
evenly distributed over \emph{all} participating nodes. 
The optimal construction of such sets was already studied in the context of
the \emph{distributed match-making} problem by Mullender and
Vitányi~\cite{MulV88,ErdFF85}.  
They show that an assignment of 
sets exists such that $\card{\cset_i} \le 2\sqrt{n}$ for all $i$, 
while for all $i,j$ $\cset_i \setjoin \cset_j \neq \emptyset$. 
They also prove a matching lower bound\footnote{%
   Note that if we somehow could construct a `uniform, randomised' selection of
   the node responsible for keeping track of the current owner of a coin, then
   using this single node as the clerk set for that coin would implement a
   distribution solution to the problem. This is studied in more detail in
   section~\ref{sec-reducing}.}.

Now suppose we do have $\faulty$ dishonest nodes. 
Using the techniques outlined above, we
arrive at the following bound.
\begin{theorem}
Double spending is deterministically prevented with fixed clerk sets of size
$2 \sqrt{n(\faulty+1)}$, when there are at most $\faulty$ dishonest nodes. 
\end{theorem}
\begin{proof}
To guarantee detection of double spending we need at least $\faulty+1$ clerks in the
intersection of any two clerk sets, hence 
\[ 
  \card{\cset_i \setjoin \cset_j} > \faulty~.
\]
One way to approach this extension is as follows. 
Cluster the $n$ nodes into groups of $\faulty+1$ 
nodes each (for simplicity assume
$\faulty+1$ exactly divides $n$).
For the resulting $\frac{n}{\faulty+1}$ so-called
supernodes $N_i$, create super clerk sets
$\scset_i$ as before. 
Now for each original node $i$, let its clerk set be the union of the nodes in
the super nodes that are a member of its super clerk set $\scset_i$. 
In other words, let $j$ be a member of super node $N_i$. 
Then
\[
  \cset_j = \bigcup_{N_k \in \scset_i} N_k~.
\]
We know $\card{\scset_i} = 2 \sqrt{\frac{n}{\faulty+1}}$, and that each 
super node
covers $\faulty+1$ nodes. Hence $\card{\cset_j} \le 2 \sqrt{n(\faulty+1)}$.
By construction, for any pair $i,j$ there is an 
$N_k \in \cset_i \setjoin \cset_j$. Hence
$\card{\cset_i \setjoin \cset_j} > \faulty$.
\qed
\end{proof}

\section{Random clerk sets}
\label{sec-random}

We now consider the case where each time a node $i$ receives a coin it
generates a different random clerk set $\cset_i$ to verify that the coin is not
being double spent\footnote{
  Actually, in this case a node can use the same randomly generated 
  clerk set throughout, \emph{provided} that $\corrupted=0$. 
  This is no longer the case when we allow small multiple
  spendings, analysed in Section~\ref{sec-multispend}.
}. 
Now suppose we have $\faulty$ dishonest nodes. 
Again we assume $\corrupted=\faulty$ (because the clerk sets are regenerated
every time a coin is received, the adversary gains no advantage if it is able
to corrupt some nodes right after system initialisation).
\begin{theorem}
\label{th-single-with-faults}
Double spending is prevented with overwhelming probability 
using random clerk sets of size at least $\sqrt{\frac{n\secpar}{\log e (1-\faulty/n)}}$.
\end{theorem}

\begin{proof}
Let $\cset_i$ be given, and randomly construct $\cset_j$. 
Let $\cssize$ be the size of the clerk sets that we aim to bound. 
$\cset_j$ does not prevent double spending if it only contains
nodes not in $\cset_i$, unless they are dishonest.
To simplify analysis, let us assume that in the random construction of the set
$\cset_j$ (and the given set $\cset_i$) we are sampling with replacement. 
This way we overestimate the probability of constructing such a bad set
(because we do not reduce the possible number of bad choices that would occur
with sampling \emph{without} replacement).
We will then show that even with this overestimation, this event will occur
with probability at most $2^{-\secpar}$.

For each member $x$ of $\cset_j$, we should either pick a node not in 
$\cset_i$ (with probability $\frac{n-\cssize}{n}$), or if we do
(with probability $\frac{\cssize}{n}$), this node should be dishonest. 
Each node in $\cset_i$ has probability $\frac{\faulty}{n}$ to be dishonest. 
Hence
\[
\Prob{x \text{ is bad}} = 
   \frac{n-\cssize}{n} + \frac{\cssize}{n} \frac{\faulty}{n}~.
\]
Then
\[
\Prob{\cset_j \text{ is bad}}
 = \Big(\Prob{x \text{ is bad}}\Big)^\cssize
 = \left(\frac{n-(1-f/n)\cssize}{n}\right)^\cssize~.
\]
With $(1-\frac{1}{x})^x < e^{-1}$, the latter can be bounded from above by
$e^{-\frac{1-\faulty/n}{n}\cssize^2}$.
We require $\Prob{\cset_j \text{ is bad}} \le 2^{-\secpar}$. This is achieved when
\[
e^{-\frac{1-\faulty/n}{n}\cssize^2} < 2^{-\secpar}~.
\]
Taking logarithms and rearranging proves the theorem.
\qed
\end{proof}
This improves the deterministic case, where we have a $\sqrt{\faulty}$ dependence on
$\faulty$.

\section{When coins get spent more often}
\label{sec-multispend}

Clearly, the problem of double spending becomes more pressing when coins
are double spent (much) more than once. We will now show that this can be
prevented with high probability with even small clerk sets. Note that
multiple double spending only helps reducing the size of the clerk sets in the
randomised case: in the deterministic case either the first double spending is
prevented straight away, or no double spending is prevented at all.

Let $\dlimit$ be the number of times a single coin is double spent by the same
node\footnote{%
  Recall that when a node spends the same coin $x$ times, then $\dlimit=x-1$.
}
We first consider the failure free case, \ie 
except for the node trying to double spend,
there are no dishonest nodes.
This case captures the situation where nodes in the clerk sets have no
incentive to collaborate with the double spender to let him get away 
undetected, and is closely related to the selfish but rational
models used in game theoretic analysis of security 
protocols (\cf ~\cite{izmalkov2005rationalsecure}).
\begin{theorem}
\label{th-multiple-no-faults}
When only the owner of a coin is dishonset,
double spending of a single coin at least $\dlimit$ times is prevented with
overwhelming probability using random clerk sets of size $\cssize$ such that
$\cssize > \frac{\sqrt{2n\secpar}}{\dlimit}+1$
(or $\cssize > \frac{n-1}{\dlimit+1}$).
\end{theorem}
\begin{proof}
Let $\cset_i$ be the set used for the verification of the coin when it is spent
for the $i$-th time. Let $q$ be the node double spending.
There are $\dlimit+1$ such sets if the coin is double spent $\dlimit$ 
times. 
If double spending is not
detected one of those $\dlimit$ times, the adversary wins. 
This happens when
$\cset_i \setjoin \cset_j$ contains at most the double spender $q$
itself, for all pairs $i,j$. 
The probability that this happens is computed as follows (where we assume
$(\dlimit+1) \cssize \le n$ or else such a collection of sets simply does not
exist).

After constructing the $i$-th set such that none of the $i$ sets (each with
$\cssize$ members) do mutually intersect except on the double spender $q$, 
there are at most $n-i(\cssize-1)$ nodes to choose
from for the $i+1$-th set, and the probability that this set does not intersect
the $i$ others except on $q$ becomes at most
$\binom{n-i(\cssize-1)}{\cssize} / \binom{n}{\cssize}$. 
Expanding binomials to their factorial representation, and cancelling
factorials in nominators and denominators, we conclude that this is less than
\[
\left(\frac{n-i(\cssize-1)}{n-\cssize+1}\right)^\cssize~.
\]
Hence
\[
\Prob{\text{double spending not detected}} \le
  \prod_{i=1}^{\dlimit} \frac{\binom{n-i(\cssize-1)}{\cssize}}{\binom{n}{\cssize}}
\le
  \prod_{i=1}^{\dlimit}  \left(\frac{n-i(\cssize-1)}{n-\cssize+1}\right)^\cssize~.
\]
Further simplification using 
$\frac{a-b}{n} \frac{a+b}{n} \le \frac{a^2}{n^2}$ shows that
this is bounded from above by
\[
\left(\frac{n-\frac{\dlimit+1}{2}(\cssize-1)}{n-\cssize+1}\right)^{\dlimit\cssize}~.
\]
We want this latter expression to be negligible, \ie less than $2^{-\secpar}$. 
Inverting fractions and taking logarithms this leads to the 
inequality
\[
\dlimit\cssize \log\left(\frac{n-\cssize+1}{n-\frac{\dlimit+1}{2}(\cssize-1)}\right)
 > \secpar~.
\]

Using $(\dlimit+1) \cssize \le n$ we see
$\frac{n-\cssize+1}{n-\frac{\dlimit+1}{2}(\cssize-1)} \le 2$.
Using this, and the fact that $\log(1+x) \ge x$ for all 
$x$ between $0$ and $1$, we have
\[
  \log\left(\frac{n-\cssize+1}{n-\frac{\dlimit+1}{2}(\cssize-1)}\right) \ge
    \left(\frac{\frac{\dlimit-1}{2}(\cssize-1)}{n-\frac{\dlimit+1}{2}\cssize}\right)
\]
Hence we require
\[
   \dlimit\cssize \left(\frac{\frac{\dlimit-1}{2}(\cssize-1)}{n-\frac{\dlimit+1}{2}\cssize}\right) 
   > \secpar
\]
Simplifying this proves the theorem.
\qed
\end{proof}
Next, we consider the case when there are at most $\faulty>1$ dishonest nodes.
\begin{theorem}
\label{th-multiple-with-faults}
Double spending of a single coin at least $\dlimit$ times is prevented with
overwhelming probability using random clerk sets of size at least
$\sqrt{\frac{n\secpar}{\log e (1-\faulty/n)\dlimit}}$.
\end{theorem}
\begin{proof}
Again, 
let there be $\dlimit+1$ sets $\cset_i$, each used for the verification of the coin
when it is spent for the $i$-th time. Let $F$ denote the set of faulty
nodes.
If double spending is not
detected one of those $\dlimit+1$ times, the adversary wins. 
This happens when 
\[
(\cset_i \setjoin \cset_j) \setminus F = \emptyset, \text{for all $i,j$}~.
\]
We are going to estimate the probability that this happens by only considering
$\cset_1 \setjoin \cset_j \setminus F = \emptyset$ for all $j \neq 1$. Then
\begin{eqnarray*}
\Prob{\text{double spending not detected}} & < &
  \left( \Prob{\cset_1 \setjoin \cset_j \setminus F = \emptyset} \right)^\dlimit \\
   & < &
    \left( \Prob{x \not\in \cset_1 \lor x \in F}^\cssize \right)^\dlimit~,
\end{eqnarray*}
where in the last step we consider arbitrary $x$ and sample with replacement.
This latter probability is, like the proof in
Theorem~\ref{th-single-with-faults} 
\[
\Prob{x \text{ is bad}} = \frac{n-\cssize}{n} + \frac{\cssize}{n} \frac{\faulty}{n}~.
\]
Proceeding similar to that proof, we obtain
$\cssize > \sqrt{\frac{n\secpar}{\log e(1-\faulty/n)\dlimit}}$.
\qed
\end{proof}
The bound appears not to be tight (in fact it is worse than
Theorem~\ref{th-multiple-no-faults} by a factor $\sqrt{\dlimit}$) because we
only estimated the probability that no clerk set intersects with the first
clerk set, thus greatly exaggerating the success of the adversary.
Simulations suggest that the size of the clerk sets $\cssize$
is indeed inversely proportional to the number of clerk sets $\dlimit$ even
when faulty nodes exist.

\section{Coin-specific clerk sets}
\label{sec-reducing}

Up till now, we have assumed that clerk sets are constructed independent of the
coin that needs to be checked. This is a restriction. In fact, we will now show
that under certain circumstances, the use of the coin identifier in the
construction of the clerk sets may help reducing the size of the clerk sets
even further. 

In previous work on digital karma~\cite{garcia2005karma} 
we investigated the design of a decentralised currency for P2P networks with
double-spending \emph{detection}. We showed the following result, given an
assignment of $\cspacesize$ nodes derived from a coin identifier $\cid$ by
\[
\cspace_{\cid}=\{ h^i(\cid) \bmod n \mid 1 \le i \le \cspacesize \}
\]
(where we ignore the possibility of collisions for the moment)
where $h$ is a random hash function.
\begin{lemma}[\cite{garcia2005karma}]
\label{lem-karma-single}
If
$\cspacesize >  \corrupted + \frac{\secpar}{\log((n-\corrupted)/(\faulty-\corrupted))}$, 
then $\cspace_{\cid}$
contains only dishonest nodes with probability less than $2^{-\secpar}$.
\end{lemma}
Note that in the proof of this result we use the fact that
the adversary controls at most $\corrupted$ nodes for which it knows
membership of a particular set $\cspace_{\cid}$; for all other $\faulty-\corrupted$
dishonest nodes membership of this set is entirely random.

Using this new approach as a starting point, we now analyse how frequent
double spending of a single coin can be prevented more efficiently. 

Clearly, when there are no dishonest nodes, the single node clerk set 
$\cset_{\cid} = \{ h(\cid) \}$ suffices to prevent double spending
(provided of course that the coin is never spent by this particular node
itself). 
This is a distributed solution because the hash function distributes the
clerk assignment uniformly over all available nodes.

Similarly, using the Lemma~\ref{lem-karma-single}, we see that using 
$\cspace_{\cid}$ as the clerk set each time coin $\cid$ is spent, double
spending is prevented with overwhelming probability as well, even
if the adversary gets to corrupt $\corrupted$ out of $\faulty$ nodes
of his own choosing. This is summarised in the following theorem.

\begin{theorem}
\label{th-single-cid}
Double spending is prevented with overwhelming probability 
using clerk sets derived from a coin identifier, of size at least
$\cspacesize > 
  \corrupted + \frac{\secpar}{\log((n-\corrupted)/(\faulty-\corrupted))}$.
\end{theorem}
But we can do even better than that if we are willing to allow a coin to be
double spent at most $\dlimit$ times.
The idea is to start with the coin-specific clerk space $\cspace_{\cid}$
of size $\cspacesize$, but to
use a smaller random subset $\cset_i \subset \cspace_{\cid}$ of size $\cssize$ as
the clerk set to use when spending the coin for the $i$-th time.

Observe that the size of the clerk space
now is more or less independent of $n$:
it only depends on the fraction of dishonest nodes.  Compared to the original
randomised clerk set case (see Theorem~\ref{th-single-with-faults}) when
setting $\corrupted=0$ we see that $\cspacesize$ increases much less rapidly with
increasing 
fraction of dishonest nodes.  Note that reducing the sample space in this
original case from $n$ to say $n'$ would improve the bound; however, the
solution would no longer be distributed because certain nodes \emph{never}
would become members of a clerk set.

\begin{theorem}
\label{th-karma-multi}
Double spending of a single coin $\cid$ at least $\dlimit$ times is
prevented with overwhelming probability using coin specific clerk spaces
of size $\cspacesize$ at least
$\corrupted + \frac{\secpar}{\log((n-\corrupted)/(\faulty-\corrupted))}$
and clerk sets of size $\cssize$ at least
$\frac{\cspacesize}{\dlimit \log e} \left( \secpar + 1 + \log (\dlimit+2) \right)$
\end{theorem}
\begin{proof}
Consider an arbitrary coin with coin identifier $\cid$.
Let $\cspacesize = |\cspace_{\cid}|$.
From Lemma~\ref{lem-karma-single} we know that if
$\cspacesize >  \corrupted + 
  \frac{\secpar+1}{\log((n-\corrupted)/(\faulty-\corrupted))}$, 
then $\cspace_{\cid}$
contains no honest nodes with negligible probability $2^{-(\secpar+1)}$.

Let this coin be double spent $\dlimit > 1$ times, and let
$\cset_i \subset \cspace_{\cid}$  be a random subset of size $\cssize$
that serves as
the clerk set to use when spending the coin for the $i$-th time.
We will show that when $\cspace_{\cid}$ contains at least one honest node $x$,
the probability that $x$ is not a member of at least two sets $\cset_i$ and 
$\cset_j$
is again at most $2^{-(\secpar+1)}$.
Multiplying these two probabilities 
we can conclude that the adversary can only succeed spending the coin
$\dlimit$ times with probability at most $2^{-\secpar}$, which proves the theorem.

We bound the probability that $x$ is not a member of at least two sets $\cset_i$
and $\cset_j$ as follows. We have
\[
\Prob{x \not\in \csset_i} = 
\frac{\cspacesize-1}{\cspacesize} \frac{\cspacesize-2}{\cspacesize-1} \cdots
  \frac{\cspacesize-\cssize}{\cspacesize-\cssize+1} = 1 - \frac{\cssize}{\cspacesize}~.
\]
Call this probability $p$. Then $q = 1-p = \frac{\cssize}{\cspacesize}$. Let
$X$ be a random variable denoting the number of sets $\cset_i$ of which $x$ is a
member. Then
\[
  \Prob{X \le 1} = p^{r+1} + \binom{\dlimit+1}{1}p^{\dlimit}q~.
\]
Assume for the moment that $\cssize > \cspacesize/2$. Then $q>p$ and hence
$\Prob{X \le 1} \le (\dlimit+2) q p^\dlimit$, which should be less than
$2^{-(\secpar+1)}$. Substituting the values for $p$ and $q$ 
and using $\frac{\cssize}{\cspacesize} \le 1$, this is achieved when
\[
  (\dlimit+2) \left( 1 - \frac{\cssize}{\cspacesize} \right)^\dlimit
   \le 2^{-(\secpar+1)}~.
\]
Using $(1-1/x)^x \le 1/e$ and taking logarithms we need
\[
  \log (\dlimit+2) - \dlimit \log e \frac{\cssize}{\cspacesize} \le - (\secpar+1)
\]
From this the theorem follows.
\qed
\end{proof}
The proof of this theorem uses a rather crude approximation of the probability
that an adversary can cheat. In fact, it is far more likely that a coin
specific clerk space contain more than one honest node, making it harder for
the adversary to avoid them in the $\dlimit$ clerk sets.

\section{Conclusions \& Further Research}
\label{sec-concl}

Interestingly, the probability of polling the central bank in the scheme of
Jarecki and Odlyzko~\cite{jarecki1997micropayment-polling} is proportional 
to the amount of the transfer, such that the number of polling messages is
constant for a given amount of credit: whether a user spends all her credit in
a few big transactions, or many micro payments does not matter. 
To get a similar property in our scheme would require us to change the size of
the clerk sets depending on the amount of the transaction (\ie the value of the
coin, if there are multi valued coins in the system), or to contact the clerk
sets only with a certain probability for each transaction. 
Further research is necessary to explore these ideas and to determine their
impact on the efficiency of double spending prevention in a decentralised,
distributed currency scheme.

The current analysis is based on a few strong assumptions.
For one thing, we assume that the network is static. To fully apply our
ideas to for instance P2P networks requires us to take dynamic node joins and
leaves into account.
Also, we assume transmitting coins is an atomic operation. Probably, the 
coin transfer protocol becomes slightly more involved when we need to handle
concurrent coin spending.
Finally, the coin transfer protocol assumes that coins can grow unbounded in
size: with every transfer of a coin, it gains another signature.
Methods to reduce the space complexity should be investigated. This is not easy
however, because the double spending prevention system depends on a more or less
correct notion of time, and aims to record who owns which coin at what time. 
Preventing nodes to warp the coins they own into the future (and thus bypassing
all double spending prevention) is not trivial.
We do note however, that clerks only need to store the coin with the longest
prefix for a particular coin identifier.

Finally, there are other interesting approaches that might be useful
to implement distributed double spending prevention. 

One approach is to try to limit the rate at which nodes can spend coins in the
first place. HashCash~\cite{back1997hashcash} could be used to do this.
In this setting, a node wishing to spend a coin is forced to spend a
non-negligible amount of work first to compute some function, \eg by finding a
collision in a moderately strong hashfunction. The receiver of the coin verifies
the function result and only accepts the coin when the result is correct.
If a lower bound on the actual time needed to compute the function is known
(and this is not always easy given the diversity of hardware platforms), 
this implies an upper bound on the amount of money a coin spent (and therefore
double spend).

\bibliography{double-spending}

\end{document}